\documentclass{article}
\usepackage{ijcai23}

\usepackage{times}
\usepackage{soul}
\usepackage{url}
\usepackage[hidelinks]{hyperref}
\usepackage[utf8]{inputenc}
\usepackage[small]{caption}
\usepackage{graphicx}
\usepackage{amsmath}
\usepackage{amsthm}
\usepackage{booktabs}
\usepackage{algorithm}
\usepackage{algorithmic}
\usepackage[switch]{lineno}
\usepackage[utf8]{inputenc}

\usepackage{enumitem}

\usepackage[utf8]{inputenc}
\usepackage{fancyhdr}

\fancypagestyle{firstpage}{
  \fancyhf{}  % Clear all header and footer fields
  \fancyfoot[C]{International Symposium on Large Language Models for Financial Services (FinLLM 2023)@IJCAI 2023 — https://finllm.github.io/workshop} 
}

\nolinenumbers

\title{FinGPT: Open-Source Financial Large Language Models}

\author{
    Hongyang Yang$^{1}$, Xiao-Yang Liu$^2$, Christina Dan Wang$^3$\thanks{Corresponding author.}
    \affiliations
    $^1$AI4Finance Foundation\thanks{AI4Finance Foundation: ai4finance.org}; \\ $^2$Columbia University; \\$^3$New York University Shanghai
    \emails
    contact@ai4finance.org 
}
\begin{document}

\maketitle

\begin{abstract}
Large language models (LLMs) have shown the potential of revolutionizing natural language processing in diverse domains, sparking great interest in finance. However, the finance domain presents unique challenges, including high temporal sensitivity, constant dynamism, and a low signal-to-noise ratio (SNR). While proprietary models like BloombergGPT have taken advantage of their unique data accumulation, such privileged access calls for an open-source alternative to democratize internet-scale financial data.

In this paper, we present an open-source large language model, FinGPT, for the finance sector. Unlike proprietary models, FinGPT takes a data-centric approach, providing researchers and practitioners with accessible and transparent resources to customize their financial LLMs (FinLLMs). We highlight the importance of an automatic data curation pipeline and the lightweight low-rank adaptation technique in building FinGPT. Furthermore, we provide fundamental tasks as building blocks for benchmarking and showcase potential applications as stepping stones for users, such as robo-advising and sentiment analysis. Through collaborative efforts within the open-source AI4Finance community, FinGPT aims to stimulate innovation, democratize FinLLMs, and unlock new opportunities in open finance. Two associated code repos are \url{https://github.com/AI4Finance-Foundation/FinGPT} and \url{https://github.com/AI4Finance-Foundation/FinNLP}
\end{abstract}

\section{Introduction}

The continual expansion and evolution of artificial intelligence have provided a fertile ground for the proliferation of LLMs \cite{NIPS2017_3f5ee243,radford2018improving,devlin2018bert,ethayarajh2019contextual,lewis2019bart,lewis2020pretrained,brown2020language,thoppilan2022lamda}, thereby effecting a transformative shift across diverse domains. This sweeping change has engendered keen interest in the potential applications of financial LLMs (FinLLMs). It is, however, evident that the acquisition of high-quality, relevant, and up-to-date data stands as a critical factor in the development of efficacious and efficient FinLLMs.

%Difficulties with LLM in Finance
%学者和读者比较想看看finance领域是否对LLMs提出了比较切实的新的，独特的挑战
%类似于，以前FinRL-meta时提的，可以是finance field requires LLMs adapt to most update information (temportal sensitive), 以及finance data普遍被认为是low SNR, even detection is more important than prediction
%LLMs may have encoded prior knowledge of anomaly events, and those events may play a key role in making investment strategy or risk control

Utilizing LLMs in the finance sector reveals intricate hurdles. Firstly, there's the issue of high temporal sensitivity. Financial data are characterized by their time-sensitive nature. Market-moving news or updates, once released, provide a narrow window of opportunity for investors to maximize their alpha (the measure of an investment's relative return). Secondly, the financial landscape is marked by high dynamism. It is in a constant state of flux due to the ceaseless flow of news, social media updates, and other market-related information. Given these constant changes, retraining LLMs frequently is not only expensive but also impractical. Lastly, financial data is often characterized by a low signal-to-noise ratio (SNR) \cite{yang2020deep}. The useful information is often hidden amongst a significant amount of irrelevant or noisy data. Extracting valuable insights from this sea of information necessitates advanced techniques.

In the proprietary sphere, models like BloombergGPT \cite{wu2023bloomberggpt} have capitalized on their exclusive access to specialized data to train a FinLLM. However, the restricted accessibility and transparency of their data collections and training protocols have accentuated the demand for an open and inclusive alternative. In response to this demand, we are witnessing a shifting trend towards democratizing internet-scale financial data in the open finance domain.

%Our work

In this paper, we address these aforementioned challenges associated with financial data and introduce FinGPT, an end-to-end open-source framework for financial large language models (FinLLMs). Adopting a data-centric approach, FinGPT underscores the crucial role of data acquisition, cleaning, and preprocessing in developing open-source FinLLMs. By championing data accessibility, FinGPT aspires to enhance research, collaboration, and innovation in finance, paving the way for open finance practices. 

Our contributions are summarized as follows:
\begin{itemize}[leftmargin=*]
    
    \item \textbf{Data-centric approach}: Recognizing the significance of data curation, FinGPT adopts a data-centric approach and implements rigorous cleaning and preprocessing methods for handling varied data formats and types.
    
    \item \textbf{End-to-end framework}: FinGPT embraces a full-stack framework with five layers:
    
    \begin{itemize}
        \item \textit{Data source layer}: Assures comprehensive market coverage, addressing the temporal sensitivity of financial data through real-time information capture.
        
        \item \textit{Data engineering layer}: Primed for real-time NLP data processing, this layer tackles the inherent challenges of high temporal sensitivity and low signal-to-noise ratio in financial data.
        
        \item \textit{LLMs layer}: Focusing on a range of fine-tuning methodologies, this layer takes care of the highly dynamic nature of financial data, ensuring the model's relevance and accuracy.
        \item \textit{Tasks Layer}: This layer is responsible for executing fundamental tasks. These tasks serve as the benchmarks for performance evaluations and cross-comparisons in the realm of FinLLMs.
        
        \item \textit{Applications layer}: Showcasing practical applications and demos, this layer highlights the potential capability of FinGPT in the finance sector.
    \end{itemize}

    \item \textbf{Democratization}: FinGPT, as an open-source framework, aims to democratize financial data and FinLLMs, uncovering untapped potentials in open finance. We envision FinGPT as a catalyst for stimulating innovation within the finance domain. FinGPT is not limited to providing technical contributions, but also cultivates an open-source ecosystem for FinLLMs, promoting real-time processing and customized adaptation for users. By nurturing a robust collaboration ecosystem within the open-source AI4Finance community, FinGPT is positioned to refine our understanding and application of FinLLMs.
\end{itemize}

\section{Related Work}

\subsection{The Raise of FinLLMs}

Large Language Models (LLMs) have been recognized as a technological breakthrough in NLP, such as GPT-3 and GPT-4 \cite{brown2020language,jiang2023mistral7b,openai_chatgpt,team2023gemini,liu2024deepseek}. They take transformer-based architectures, demonstrating impressive performance across various text-generation tasks. As an offshoot of the GPT family developed by OpenAI, ChatGPT was designed to produce human-like texts based on input prompts. It has shown significant utility in diverse applications, from drafting emails to writing code and even in creating art content.

LLMs have been applied to various tasks within the financial sector \cite{dredze2016twitter,araci2019finbert,bao2021plato,delucia2022bernice}, from predictive modeling to generating insightful narratives from raw financial data. Recent literature has focused on using these models for financial text analysis, given the abundance of text data in this field, such as news articles, earnings call transcripts, and social media posts.

The first example of FinLLMs is BloombergGPT \cite{wu2023bloomberggpt}, which was trained on a mixed dataset of financial and general data sources. Despite its impressive capabilities, access limitations exist, and the prohibitive training cost has motivated the need for low-cost domain adaptation.

Our FinGPT responds to the aforementioned hurdles, presenting an open-source FinLLM. It employs Reinforcement Learning from Human Feedback (RLHF) to understand and adapt to individual preferences, paving the way for personalized financial assistants. We aim to combine the strengths of general LLMs like ChatGPT with financial adaptation, exploiting LLM's capability in open finance.

\subsection{Why Open-Source FinLLMs?}
AI4Finance Foundation\footnote{\url{https://ai4finance.org}.~The AI4Finance Foundation is a U.S.-registered 501(c)(3) nonprofit public charity focused on promoting open scientific research in financial AI, building open-source infrastructure, and supporting a global community of researchers through shared datasets, benchmarks, and educational programs.}  is a non-profit, open-source organization that integrates Artificial Intelligence (AI) and financial applications. With a proven track record of nurturing an innovative ecosystem of FinTech tools, such as FinRL \cite{yang2020deep} and FinRobot \cite{yang2024finrobot}, the foundation is poised to accelerate the evolution of FinLLMs. Steadfast commitment and cutting-edge contributions may pave the way for AI's transformative applications in open finance.

\begin{itemize}[leftmargin=*]
    \item Advancing equal opportunities via democratizing FinLLMs: Adopting an open-source methodology promotes universal access to state-of-the-art technology, adhering to the ethos of democratizing FinLLMs. 
    \item Cultivating transparency and trust: Open-source FinLLMs offer a comprehensive overview of their foundational codebase, bolstering transparency and trust. 
    \item Accelerating research and innovation: The open-source model fuels progress in research and development within the AI domain. It allows researchers to leverage existing models, thus nurturing a faster progression of innovation and scientific discovery.
    \item Enhancing education: Open-source FinLLMs serve as robust educational tools, presenting students with the prospect of exploring the complexities of FinLLMs through direct engagement with fully operational models.
    \item Upgrade foundation infrastructure for financial text data by community collaboration: This collaborative participation bolsters the model's long-term durability and effectiveness.
\end{itemize}

\section{Overview of FinGPT: An Open-Source Framework for FinLLMs}

\begin{figure*}
\centering
\includegraphics[scale = 0.228]{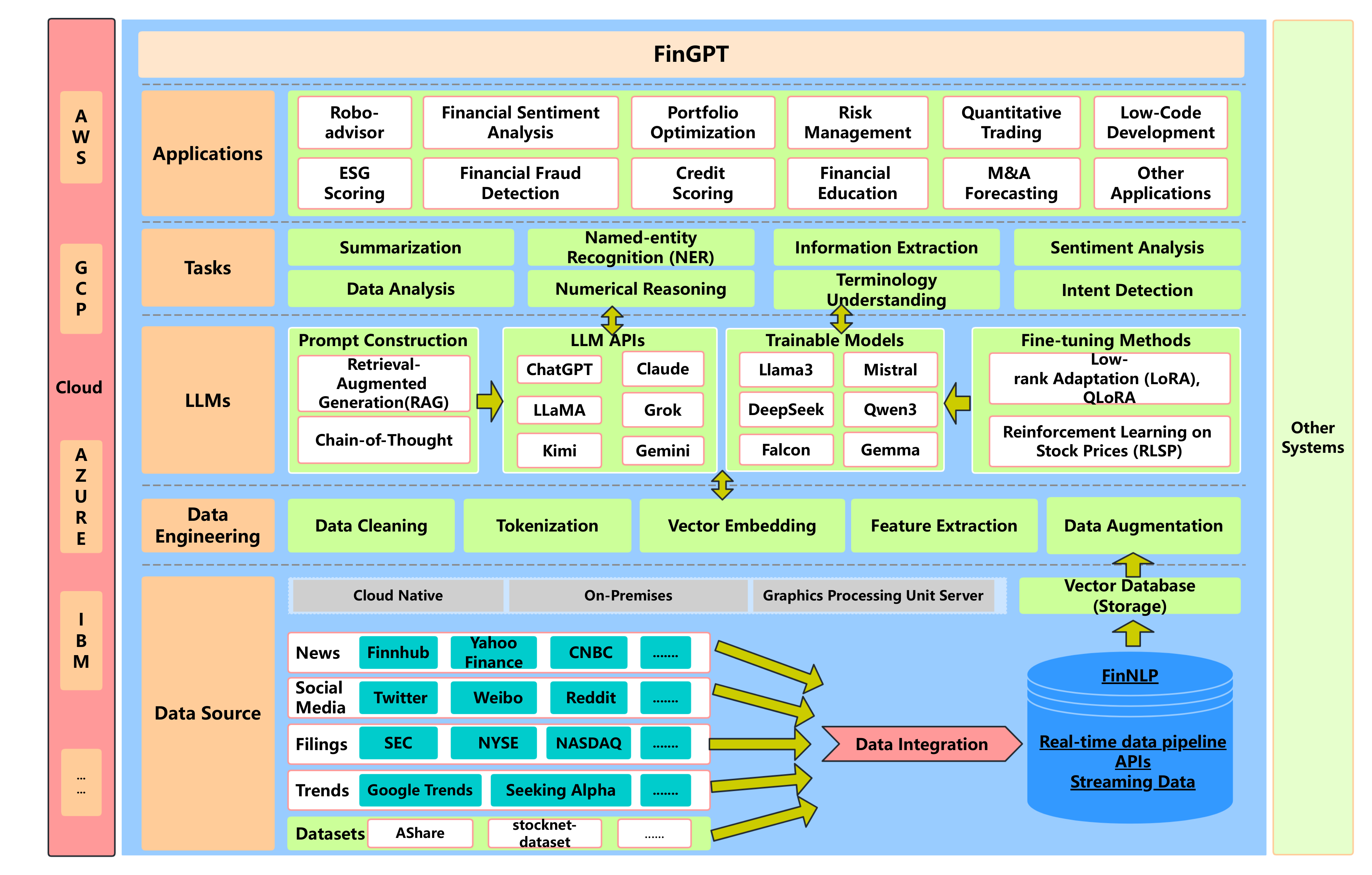}
\caption{Overall framework of FinGPT.}
\label{fig:framework}
\vspace{-1mm}
\end{figure*}

FinGPT represents an innovative open-source framework designed specifically for FinLLMs. As delineated in Fig. \ref{fig:framework}, FinGPT consists of four components: Data Source, Data Engineering, LLMs, and Applications. Each plays a crucial role in maintaining the functionality and adaptability of FinGPT.
\begin{itemize}[leftmargin=*]
\item \textbf{Data source layer}: The starting point is the Data Source Layer, which orchestrates the acquisition of extensive financial data from a wide array of online sources. This layer ensures comprehensive market coverage by integrating data from news websites, social media platforms, financial statements, market trends, and more. The goal is to capture the nuance of the market, thereby addressing the inherent temporal sensitivity.

\item \textbf{Data engineering layer}: This layer focuses on the real-time processing of text data to tackle the challenges of \textit{high temporal sensitivity} and \textit{low signal-to-noise ratio} inherent in financial data. It incorporates state-of-the-art NLP techniques to filter noise and highlight the most salient pieces of information. 

\item \textbf{LLMs layer}: Lying at the heart, it encompasses various fine-tuning methodologies, prioritizing lightweight adaptation, to keep the model updated and pertinent. By maintaining an updated model, FinGPT can take care of the highly dynamic nature of financial data, ensuring its responses are in sync with the current financial climate.

\item \textbf{Tasks layer}: The tasks layer is designed to provide building blocks. This layer serves a dual purpose: first, it executes a variety of fundamental tasks that are crucial in the FinLLMs landscape, such as sentiment analysis, content summarization, and numerical reasoning. Second, it establishes a standardized set of metrics and attributes. These standardized elements act not only as indicators but also as benchmarks, facilitating both performance evaluation and comparative analysis in the domain of FinLLMs.

\item \textbf{Application layer}: The final component of FinGPT is the Applications Layer, designed to demonstrate the practical applicability of FinGPT. It offers hands-on tutorials and demo applications for financial tasks, including robo-advisory services and sentiment analysis. These practical demos not only serve as a guide to potential users but also underscore the transformative potential of FinLLMs.

\end{itemize}

\subsection{Data Sources}

The first stage of FinGPT involves the collection of extensive financial data from a wide array of online sources. These include, but are not limited to:
\begin{itemize}[leftmargin=*]
\item \textbf{Financial news:} Websites such as Reuters, CNBC, Yahoo Finance, among others, are rich sources of financial news and market updates. These sites provide valuable information on market trends, company earnings, macroeconomic indicators, and other financial events.
\item \textbf{Social media}: Platforms such as Twitter, Facebook, Reddit, Weibo, and others, offer a wealth of information in terms of public sentiment, trending topics, and immediate reactions to financial news and events.
\item \textbf{Filings}: Websites of financial regulatory authorities, such as the SEC in the United States, offer access to company filings. These filings include annual reports, quarterly earnings, insider trading reports, and other important company-specific information. Official websites of stock exchanges (NYSE, NASDAQ, Shanghai Stock Exchange, etc.) provide crucial data on stock prices, trading volumes, company listings, historical data, and other related information.
\item \textbf{Trends}: Websites like Seeking Alpha, Google Trends, and other finance blogs and forums provide access to analysts' opinions, market predictions, the movement of specific securities or market segments and investment advice.
\item \textbf{Academic datasets}: Research-based datasets that offer curated and verified information for financial analysis.

\end{itemize}

To harness the wealth of information from these diverse sources, FinGPT incorporates data acquisition tools capable of scraping structured and unstructured data, including APIs, web scraping tools, and direct database access where available. Moreover, the system is designed to respect the terms of service of these platforms, ensuring data collection is ethical and legal.

\textbf{Data APIs}: In the FinGPT framework, APIs are used not only for initial data collection but also for real-time data updates, ensuring the model is trained on the most current data. Additionally, error handling and rate-limiting strategies are implemented to respect API usage limits and avoid disruptions in the data flow.

\subsection{Real-Time Data Curation Pipeline for Financial NLP}

Financial markets operate in real-time and are highly sensitive to news and sentiment. Prices of securities can change rapidly in response to new information, and delays in processing that information can result in missed opportunities or increased risk. As a result, real-time processing is essential in financial NLP. 

The primary challenge with a real-time NLP pipeline is managing and processing the continuous inflow of data efficiently. The first step in the pipeline is to set up a system to ingest data in real-time. This data could be streaming from our data source APIs.  Below are the steps to design a real-time NLP pipeline for data ingestion.

\textbf{Data cleaning}: Real-time data can be noisy and inconsistent. Therefore, real-time data cleaning involves removing irrelevant data, handling missing values, text normalization (like lowercasing), and error corrections.

\textbf{Tokenization}: In real-time applications, tokenization has to be performed on the fly. This involves breaking down the stream of text into smaller units or tokens.

\textbf{Vector embedding}: FinGPT encodes curated financial text into dense semantic vectors using domain-adapted embedding models. The embedding process incorporates entity-aware representations (tickers, ratios, events) and temporal metadata, allowing the system to capture fine-grained financial meaning. All embeddings are indexed in a vector database for low-latency retrieval, supporting RAG, event clustering, and market-aligned RLSP training.

\textbf{Feature extraction}: Feature extraction involves transforming raw data into an input that can be understood by ML models. In real-time systems, this often needs to be a fast and efficient process. Techniques such as TF-IDF, Bag of Words, or embedding vectors like Word2Vec can be used.

\textbf{Data augmentation}: In the dynamic landscape of financial markets, enhancing the variety and volume of training data is crucial for building robust NLP models. Data augmentation strategies will be employed to generate synthetic data that can mimic the characteristics of actual financial data.

\subsection{Large Language Models (LLMs)}

Once the data has been properly prepared, it is used with LLMs to generate insightful financial analyses. The LLM layer includes:
\begin{itemize}[leftmargin=*]
    \item \textbf{LLM APIs}: Established LLM APIs offer foundational language capabilities that serve as the base for further model development and customization.
    \item \textbf{Trainable models}: Users can fine-tune FinGPT's trainable models on private data for personalized financial applications, ensuring relevance and accuracy in specific use cases.
    \item \textbf{Fine-tuning methods}: FinGPT supports various fine-tuning methodologies, facilitating its adaptation into personalized robo-advisors efficiently and effectively.
    \item \textbf{Prompt Engineering}: Prompt Engineering is crucial for optimizing input queries to LLMs, enhancing the extraction of accurate financial information. This iterative process requires careful crafting of prompts for nuanced responses, necessitating a deep understanding of both finance and language model characteristics.

\end{itemize}

\noindent \textbf{Why lightweight fine-tuning LLMs for finance?}

Fine-tuning or Instruction tuning of pre-existing LLMs for finance, as described in \cite{ouyang2022training}, presents a cost-efficient and time-saving alternative to the expensive and lengthy process of retraining models from scratch.

BloombergGPT \cite{wu2023bloomberggpt}, though remarkable in its finance-specific capabilities, comes with an intensive computational requirement. It used approximately 1.3 million GPU hours for training, which, when calculated using AWS cloud's \$2.3 rate, translates to a staggering cost of around \$3 million per training. In contrast to the high computational cost of models like BloombergGPT \cite{wu2023bloomberggpt}, FinGPT presents a more accessible solution by focusing on the lightweight adaptation of top open-source LLMs. The cost of adaptation falls significantly, estimated at around \$300 per fine-tuning.

This approach ensures timely updates and adaptability, essential in the dynamic financial domain. Being open-source, FinGPT not only promotes transparency but also allows user customization, catering to the rising trend of personalized financial advisory services. Ultimately, FinGPT's cost-effective, flexible framework holds the potential to democratize financial language modeling and foster user-focused financial services.

\subsubsection{Fine-tuning via Low-rank Adaptation (LoRA)}

In FinGPT, we fine-tune a pre-trained LLM utilizing a financial dataset. It's well recognized that high-quality labeled data is a pivotal determinant for many successful LLMs, including ChatGPT. However, acquiring such top-notch labeled data often proves costly in terms of time and resources and generally requires the expertise of finance professionals.

When the application of LLMs is envisioned for the scrutiny of financial texts and facilitation of quantitative trading strategies, it is imperative to contemplate the utilization of the intrinsic labeling mechanisms available within the financial marketplace. In light of this, FinGPT adopts the percentage of relative stock price changes corresponding to individual news articles as output labels. By assigning predetermined thresholds, these continuous labels are categorized into three discrete sentiment classes: positive, negative, and neutral.

%If our objective is to employ LLMs for analyzing financial-related text data and assisting in quantitative trading, it seems sensible to leverage the market's inherent labeling capacity. Consequently, we use the relative stock price change percentage for each news item as the output label. We establish thresholds to divide these labels into three categories—positive, negative, and neutral—based on the sentiment of the news item.

%In a corresponding step, during the prompt engineering process, we also prompt the model to select one from the positive, negative, and neutral outputs. This strategy ensures optimal utilization of the pre-trained information. By deploying the Low-Rank Adaptation (LoRA) of LLMs \cite{hu2021lora,dettmers2023qlora}, along with its quantized variant, QLoRA \cite{dettmers2023qlora}, we manage to reduce the number of trainable parameters from 6.17 billion to a mere 3.67 million.

Simultaneously, during the prompt engineering phase, the model is meticulously instructed to elect one among the three sentiment classes as its output. This meticulous approach ensures that the information gleaned during pre-training is maximally exploited, fostering the generation of insightful and reliable predictions on financial sentiment. The implementation of Low-Rank Adaptation (LoRA) for LLMs \cite{hu2021lora,dettmers2023qlora}, along with its quantized variant, QLoRA \cite{dettmers2023qlora}, significantly streamlines the model by reducing the count of trainable parameters from an overwhelming 6.17 billion to a manageable 3.67 million.

\subsubsection{Fine-tuning via Reinforcement Learning on Stock Prices (RLSP)}

Similarly, we can substitute Reinforcement Learning on Stock Prices (RLSP) for Reinforcement Learning on Human feedback, as utilized by ChatGPT. The reasoning behind this substitution is that stock prices offer a quantifiable, objective metric that reflects market sentiment in response to news and events. This makes it a robust, real-time feedback mechanism for training our model.

Reinforcement Learning (RL) allows the model to learn through interaction with the environment and receiving feedback. In the case of RLSP, the environment is the stock market, and the feedback comes in the form of stock price changes. This approach permits FinGPT to refine its understanding and interpretation of financial texts, improving its ability to predict market responses to various financial events. By associating news sentiment with the subsequent performance of the related stocks, RLSP provides an effective way to fine-tune FinGPT. In essence, RLSP allows the model to infer the market’s response to different news events and adjust its understanding and predictions accordingly.

Therefore, the integration of RLSP into the fine-tuning process of FinGPT provides a powerful tool for improving the model's financial market understanding and predictive accuracy. By using actual stock price movements as feedback, we are directly harnessing the wisdom of the market to make our model more effective.

\subsubsection{Retrieval Augmented Generation (RAG)} Retrieval-augmented generation (RAG) is a pivotal technique incorporated within FinGPT \cite{zhang2023fingptrag}, as it seamlessly amalgamates the prowess of both context retrieval mechanisms and Large Language Models (LLMs) to optimize language generation tasks. This meticulous process ensures that the LLMs are not generating content in a vacuum but are rather informed and nuanced in their output, drawing from a rich tapestry of context provided by the retrieved documents. These documents, working in tandem with the input prompt, steer the LLMs effectively towards crafting responses that are not only accurate but deeply ingrained in the relevant context, thereby increasing the utility and reliability of the generated text.

\subsection{Fundamental Tasks}

FinGPT serves as a versatile tool in the financial sector, providing valuable assistance to both professionals and individuals by effectively filtering and analyzing information. The model excels in the following fundamental tasks:

\begin{itemize}[leftmargin=*]
    \item \textbf{Summarization}: FinGPT can efficiently condense lengthy financial documents into concise summaries, preserving the crucial information and insights. This function is invaluable for quickly understanding the essence of comprehensive reports, news articles, or financial statements without going through the entire content.

    \item \textbf{Named-entity recognition (NER)}: The model is adept at identifying and classifying named entities within the text, such as company names, stock tickers, monetary values, and percentages. This ability is crucial for extracting specific data points from unstructured text, facilitating more structured and informed analysis.

    \item \textbf{Information extraction}: FinGPT can meticulously extract relevant information from various sources, providing users with valuable insights. This capability is crucial for decision-making processes, as it sifts through the noise to highlight essential data and trends.

    \item \textbf{Sentiment analysis}: Sentiment Analysis is pivotal as a fundamental task due to its dual application in both identifying market sentiment, namely financial sentiment analysis and being utilized within robo-advisory platforms to discern client emotions during product recommendations. 

    \item \textbf{Data analysis}: FinGPT can process and analyze vast datasets, identifying patterns, anomalies, and significant changes in the data. This feature supports data-driven decision-making, offering a clearer understanding of market dynamics and financial performance.

    \item \textbf{Numerical reasoning}: The model can perform calculations and numerical analysis based on the data provided in the text, supporting users in evaluating financial metrics, making projections, and assessing risks effectively.

    \item \textbf{Terminology understanding}: FinGPT is proficient in understanding and interpreting complex financial terminology and jargon, making it a valuable assistant for both seasoned professionals and individuals new to the financial sector.

    \item \textbf{Intent detection}: The model can accurately identify the user’s intent behind a query, facilitating more effective and relevant responses. This feature is particularly useful in developing intuitive and user-friendly financial advisory applications and services.
\end{itemize}

Various open-source datasets serve as benchmarks, effectively engaging in a multitude of fundamental tasks. Examples include BloombergGPT\cite{wu2023bloomberggpt}, which utilizes a curated selection of financial datasets derived from the FLUE benchmark\cite{shah2022flue}. These datasets are employed for a spectrum of essential tasks such as Sentiment Analysis and NER. Other noteworthy datasets include FinRED \cite{finred2022}, instrumental for information extraction tasks, FINQA \cite{chen2021finqa} for numerical reasoning assessments, and FinRAD \cite{finread2021}, which is crucial for understanding and identifying financial terms.

\subsection{Potential Applications}

FinGPT may find wide applications in financial services, aiding professionals and individuals as a powerful information filter. The potential applications include:
\begin{itemize}[leftmargin=*]

    \item \textbf{Financial sentiment analysis}:  Evaluating sentiments across different financial platforms for insightful investment guidance.

\item \textbf{Robo-advisor}: The Robo-advisor function within FinLLMs plays a pivotal role in providing personalized financial advice, minimizing the necessity for continual human consultations. 

    \item \textbf{Quantitative trading}: Producing trading signals for informed trading decisions.
    \item \textbf{Portfolio optimization}: Utilizing numerous economic indicators and investor profiles for optimal investment portfolio construction.
    \item \textbf{Credit scoring}: Predicting creditworthiness from financial data to aid lending decisions.
    \item \textbf{Mergers and acquisitions (M\&A) forecasting}: Predicting potential M\&A activities by analyzing financial data and company profiles, helping investors anticipate market movements.
    \item \textbf{ESG (Environmental, Social, Governance) scoring}: Evaluating companies' ESG scores by analyzing public reports and news articles.

    \item \textbf{Risk management:} Formulating effective risk strategies by analyzing various risk factors.

    \item \textbf{Fraud detection}: Identifying potential fraudulent transaction patterns for enhanced financial security.
    
    \item \textbf{Automating KYC Processes:} FinGPT can streamline KYC procedures by analyzing documents for identity validation, cross-checking information against databases, and detecting inconsistencies. It can also interpret complex legal documents using its NLP capabilities.

    \item \textbf{Enhancing Anti-Money Laundering (AML) Measures:} FinGPT can be a valuable tool in ML operations. It can be used to analyze the flow of funds, identify suspicious patterns, and highlight transactions that require further investigation.

    \item \textbf{Low-code development}: Facilitating software creation through user-friendly interfaces, reducing reliance on traditional programming.

    \item \textbf{Financial education}: Serving as an AI tutor simplifying complex financial concepts for better financial literacy.
    
    \end{itemize}

By linking these distinct yet interconnected components, FinGPT provides a holistic and accessible solution for leveraging AI in finance, facilitating research, innovation, and practical applications in the financial industry.

\section{Data-Centric Approach for FinLLMs}
For financial large language models (FinLLMs), a successful strategy is not solely based on the capability of the model
architecture but is equally reliant on the training data. Our data-centric approach prioritizes collecting, preparing, and
processing financial data.

Financial data comes from a variety of sources, with unique characteristics. We delve into the specifics of different financial data sources, such as financial news, company fillings and announcements, social media Discussions, and trends.

\subsection{Financial News} 

Financial news carries vital information about the world economy, specific industries, and individual companies. This data source typically features:
\begin{itemize}[leftmargin=*]
    \item \textbf{Timeliness:} Financial news reports are timely and up-to-date, often capturing the most recent developments in the financial world. 
    \item \textbf{Dynamism:} The information contained in financial news is dynamic, changing rapidly in response to evolving economic conditions and market sentiment. 
    \item \textbf{Influence:} Financial news has a significant impact on financial markets, influencing traders' decisions and potentially leading to dramatic market movements.
\end{itemize}

\subsection{Company Filings and Announcements} 

Company filings and announcements are official documents that corporations submit to regulatory bodies, providing insight into a company's financial health and strategic direction. They feature:
\begin{itemize}[leftmargin=*]
    \item \textbf{Granularity}: These documents offer granular information about a company's financial status, including assets, liabilities, revenue, and profitability.
    \item \textbf{Reliability}: Company fillings contain reliable and verified data vetted by regulatory bodies. 
    \item \textbf{Periodicity}: Company fillings are periodic, usually submitted on a quarterly or annual basis, offering regular snapshots of a company's financial situation.
    \item \textbf{Impactfulness}: Company announcements often have substantial impacts on the market, influencing stock prices and investor sentiment.
\end{itemize}

\subsection{Social Media Discussions} 

Social media discussions related to finance will reflect public sentiment towards specific stocks, sectors, or the overall market. These discussions tend to exhibit:
\begin{itemize}[leftmargin=*]
    \item \textbf{Variability:} Social media discussions vary widely in tone, content, and quality, making them rich, albeit complex, sources of information.
    \item \textbf{Real-time sentiment:} These platforms often capture real-time market sentiment, enabling the detection of trends and shifts in public opinion.
    \item \textbf{Volatility:} Sentiments expressed on social media can be highly volatile, changing rapidly in response to news events or market movements.
\end{itemize}

\subsection{Trends} 

Trends often observable through websites like Seeking Alpha, Google Trends, and other finance-oriented blogs and forums, offer critical insights into market movements and investment strategies. They feature:
\begin{itemize}[leftmargin=*]
    \item \textbf{Analyst perspectives:} These platforms provide access to market predictions and investment advice from seasoned financial analysts and experts.
    \item \textbf{Market sentiment:} The discourse on these platforms can reflect the collective sentiment about specific securities, sectors, or the overall market, providing valuable insights into the prevailing market mood.
    \item \textbf{Broad coverage:} Trends data spans diverse securities and market segments, offering comprehensive market coverage.
\end{itemize}

Each of these data sources provides unique insights into the financial world. By integrating these diverse data types, FinGPT can facilitate a comprehensive understanding of financial markets and enable effective financial decision-making.

\section{Experiments: Financial Sentiment Analysis}

In this section, we evaluate the sentiment analysis capability of FinGPT. This experiment demonstrates the effectiveness of FinGPT’s data-centric design and lightweight adaptation methodology in real-world financial text classification.

\subsection{Dataset}

We utilize a large-scale financial news sentiment dataset curated through the FinGPT real-time data pipeline. The dataset contains:

\begin{itemize}
    \item Over \textbf{620,000} cleaned financial news headlines;
    \item Sources including \textit{CNBC, Reuters, Yahoo Finance, MarketWatch, etc.}, collected through the FinNLP pipeline;
    \item Time span from \textbf{2016--2024};
    \item Market-driven labels generated using short-term price movement:
\end{itemize}

\[
\text{label} =
\begin{cases}
\text{Positive}, & r > \theta_p \\
\text{Negative}, & r < -\theta_n \\
\text{Neutral}, & |r| \le \theta
\end{cases}
\]

where $r$ denotes the stock’s percentage price change following the news.  
This “self-labeled” approach aligns sentiment with true market reactions and avoids costly manual annotation.

\subsection{Model and Training Setup}

We adopt a lightweight two-stage adaptation process.

\subsubsection{LoRA-based Supervised Fine-Tuning}

We fine-tune the pretrained \texttt{Llama-3.1-8B-Instruct} model using
Low-Rank Adaptation (LoRA). Under a standard configuration of rank
$r=8$ and scaling factor $\alpha=16$, the total number of trainable
parameters introduced by LoRA is approximately \textbf{8.3M}, which is
well below 0.1\% of the original 8B-parameter model.

The fine-tuning configuration is as follows:

\begin{itemize}
    \item Trainable parameters: \textbf{8.3M};
    \item LoRA rank: $r = 8$, scaling factor $\alpha = 16$;
    \item Batch size: $64$;
    \item Learning rate: $2 \times 10^{-4}$;
    \item Training epochs: $3$.
\end{itemize}

LoRA enables FinGPT to acquire domain-specific sentiment classification ability efficiently.

\subsubsection{Reinforcement Learning on Stock Prices (RLSP)}

To align the model with real market behavior, we further apply RLSP, where the environment is the financial market and the reward is the stock price’s post-news reaction.

\[
R = f(\Delta p)
\]

This aligns the sentiment output with actual financial outcomes and enhances generalization.

\subsection{Baselines}

We compare FinGPT against standard financial NLP baselines:

\begin{itemize}
    \item FinBERT~\cite{araci2019finbert};
    \item BloombergGPT~\cite{wu2023bloomberggpt};
    \item ChatGPT (zero-shot) ~\cite{openai_chatgpt};
    \item Llama3.1-8B (zero-shot) ~\cite{grattafiori2024llama}.
\end{itemize}

\subsection{Evaluation Metrics}

We evaluate performance using the following metrics:

\begin{itemize}
    \item Accuracy;
    \item Precision, Recall, and F1-score for each class;
    \item Macro-F1 (to mitigate class imbalance);
    \item AUC for binary (positive/negative) subsets.
\end{itemize}

\subsection{Results}

\subsubsection{Overall Performance}

\begin{table*}[h]
\centering
\caption{Sentiment Classification Performance}
\begin{center}
\begin{tabular}{lccccc}
\toprule
\textbf{Model} & \textbf{Acc.} & \textbf{Macro-F1} & \textbf{Pos-F1} & \textbf{Neg-F1} & \textbf{Neu-F1} \\
\midrule
ChatGPT (0-shot) & 63.4 & 61.7 & 64.0 & 59.1 & 62.0 \\
Llama3.1-8B (0-shot) & 57.9 & 54.4 & 56.1 & 53.2 & 54.0 \\
FinBERT & 71.2 & 69.9 & 73.0 & 69.1 & 67.5 \\
\textbf{FinGPT (LoRA-SFT)} & \textbf{78.8} & \textbf{77.3} & \textbf{79.6} & \textbf{76.8} & \textbf{75.4} \\
\textbf{FinGPT (SFT+RLSP)} & \textbf{82.1} & \textbf{80.9} & \textbf{83.4} & \textbf{81.5} & \textbf{77.8} \\
\bottomrule
\end{tabular}
\end{center}
\end{table*}

FinGPT outperforms all baselines significantly, demonstrating the benefits of data-centric labeling and RLSP reinforcement alignment.

\subsubsection{Ablation Study}

\begin{table}[h]
\centering
\caption{Ablation on LoRA and RLSP}
\begin{tabular}{lc}
\toprule
\textbf{Configuration} & \textbf{Macro-F1} \\
\midrule
Base Llama3 & 54.4 \\
+ LoRA SFT & 77.3 \\
+ RLSP & \textbf{80.9} \\
\bottomrule
\end{tabular}
\end{table}

LoRA performs most of the heavy lifting, while RLSP further improves market alignment.

\subsubsection{Case Study}

We illustrate the model's financial reasoning capability using the following headline:

\begin{quote}
\textit{``Tesla cuts prices again in China as EV competition intensifies.''}
\end{quote}

\begin{itemize}
    \item \textbf{Human Annotation}: Negative (Price reductions are commonly interpreted as a sign of weakening pricing power and intensified competitive pressure, both of which imply potential margin compression and typically induce negative investor sentiment.)
    \item \textbf{Base Llama3}: Neutral (The model captures the surface-level wording but fails to infer the underlying financial implications of price competition.)
    \item \textbf{FinGPT (SFT)}: Negative
    \item \textbf{FinGPT (RLSP)}: Negative (\textit{with stronger alignment to the subsequent price reaction})
\end{itemize}

This case highlights FinGPT's ability to incorporate domain-specific financial reasoning and to produce sentiment predictions that are more consistent with market-impactful interpretations.

\subsection{Discussion}

Key observations:

\begin{itemize}
    \item Market-driven labels (self-labeled data) strongly improve real-world applicability;
    \item LoRA reduces adaptation cost by $\sim\!1000\times$ compared to full fine-tuning;
    \item RLSP incorporates financial market feedback, distinguishing FinGPT from traditional supervised models.
\end{itemize}

This experiment confirms that FinGPT provides a scalable and effective foundation for financial sentiment analysis.

\section{Conclusion}

In conclusion, the transformative integration of large language models (LLMs) into the financial sector brings unique complexities and vast opportunities. Navigating challenges such as high temporal sensitivity, dynamic financial landscape, and a low signal-to-noise ratio in financial data calls for efficient solutions. FinGPT responds innovatively by leveraging pre-existing LLMs and fine-tuning them to specific financial applications. This approach significantly reduces adaptation costs and computational requirements compared to models like BloombergGPT, offering a more accessible, flexible, and cost-effective solution for financial language modeling. Thus, it enables consistent updates to ensure model accuracy and relevance, a critical aspect in the dynamic and time-sensitive world of finance.

\section{Future Work}

Future development of FinLLMs will focus on establishing open, industry-level standards for financial large language models. This includes advancing parameter-efficient fine-tuning methods such as LoRA and QLoRA to support low-cost, domain-specific customization across diverse financial institutions. Furthermore, FinLLMs will continue to expand its unified data curation pipeline, promoting high-quality, standardized financial datasets to streamline training and evaluation. By integrating open-source tooling, reproducible benchmarks, and transparent workflows, FinLLMs aims to provide a foundation for reliable, scalable, and interoperable financial AI systems.

\textbf{Disclaimer: We are sharing codes for academic purposes under the MIT education license. Nothing herein is financial advice, and NOT a recommendation to trade real money. Please use common sense and always first consult a professional before trading or investing.}

\bibliographystyle{named}
\bibliography{ref}

@article{brown2020language,
  title={Language models are few-shot learners},
  author={Brown, Tom and Mann, Benjamin and Ryder, Nick and Subbiah, Melanie and Kaplan, Jared D and Dhariwal, Prafulla and Neelakantan, Arvind and Shyam, Pranav and Sastry, Girish and Askell, Amanda and others},
  journal={Advances in Neural Information Processing Systems},
  volume={33},
  pages={1877--1901},
  year={2020}
}

@article{wu2023bloomberggpt,
  title={{BloombergGPT}: A large language model for finance},
  author={Wu, Shijie and Irsoy, Ozan and Lu, Steven and Dabravolski, Vadim and Dredze, Mark and Gehrmann, Sebastian and Kambadur, Prabhanjan and Rosenberg, David and Mann, Gideon},
  journal={arXiv preprint arXiv:2303.17564},
  year={2023}
}

@article{hu2021lora,
  title={{LoRA}: Low-rank adaptation of large language models},
  author={Hu, Edward J and Shen, Yelong and Wallis, Phillip and Allen-Zhu, Zeyuan and Li, Yuanzhi and Wang, Shean and Wang, Lu and Chen, Weizhu},
  journal={International Conference on Learning Representations},
  year={2021}
}

@article{dettmers2023qlora,
  title={{QLoRA}: Efficient Finetuning of Quantized {LLMs}},
  author={Dettmers, Tim and Pagnoni, Artidoro and Holtzman, Ari and Zettlemoyer, Luke},
  journal={arXiv preprint arXiv:2305.14314},
  year={2023}
}

@article{araci2019finbert,
  title={Finbert: Financial sentiment analysis with pre-trained language models},
  author={Araci, Dogu},
  journal={arXiv preprint arXiv:1908.10063},
  year={2019}
}

@article{bao2021plato,
  title={Plato-xl: Exploring the large-scale pre-training of dialogue generation},
  author={Bao, Siqi and He, Huang and Wang, Fan and Wu, Hua and Wang, Haifeng and Wu, Wenquan and Wu, Zhihua and Guo, Zhen and Lu, Hua and Huang, Xinxian and others},
  journal={arXiv preprint arXiv:2109.09519},
  year={2021}
}

@inproceedings{delucia2022bernice,
  title={Bernice: a multilingual pre-trained encoder for {Twitter}},
  author={DeLucia, Alexandra and Wu, Shijie and Mueller, Aaron and Aguirre, Carlos and Resnik, Philip and Dredze, Mark},
  booktitle={Proceedings of the Conference on Empirical Methods in Natural Language Processing},
  pages={6191--6205},
  year={2022}
}

@inproceedings{dredze2016twitter,
  title={How twitter is changing the nature of financial news discovery},
  author={Dredze, Mark and Kambadur, Prabhanjan and Kazantsev, Gary and Mann, Gideon and Osborne, Miles},
  booktitle={Proceedings of the second International Workshop on Data Science for Macro-modeling},
  pages={1--5},
  year={2016}
}

@inproceedings{lewis2020pretrained,
  title={Pretrained language models for biomedical and clinical tasks: understanding and extending the state-of-the-art},
  author={Lewis, Patrick and Ott, Myle and Du, Jingfei and Stoyanov, Veselin},
  booktitle={Proceedings of the 3rd Clinical Natural Language Processing Workshop},
  pages={146--157},
  year={2020}
}

@article{devlin2018bert,
  title={Bert: Pre-training of deep bidirectional transformers for language understanding},
  author={Devlin, Jacob and Chang, Ming-Wei and Lee, Kenton and Toutanova, Kristina},
  journal={arXiv preprint arXiv:1810.04805},
  year={2018}
}

@article{ethayarajh2019contextual,
  title={How contextual are contextualized word representations? comparing the geometry of BERT, ELMo, and GPT-2 embeddings},
  author={Ethayarajh, Kawin},
  journal={arXiv preprint arXiv:1909.00512},
  year={2019}
}

@inproceedings{NIPS2017_3f5ee243,
 author = {Vaswani, Ashish and Shazeer, Noam and Parmar, Niki and Uszkoreit, Jakob and Jones, Llion and Gomez, Aidan N and Kaiser, \L ukasz and Polosukhin, Illia},
 booktitle = {Advances in Neural Information Processing Systems},
 publisher = {Curran Associates, Inc.},
 title = {Attention is All you Need},
 url = {https://proceedings.neurips.cc/paper_files/paper/2017/file/3f5ee243547dee91fbd053c1c4a845aa-Paper.pdf},
 volume = {30},
 year = {2017}
}

@article{radford2018improving,
  title={Improving language understanding by generative pre-training},
  author={Radford, Alec and Narasimhan, Karthik and Salimans, Tim and Sutskever, Ilya and others},
  journal={OpenAI},
  year={2018},
  publisher={OpenAI}
}

@article{lewis2019bart,
  title={Bart: Denoising sequence-to-sequence pre-training for natural language generation, translation, and comprehension},
  author={Lewis, Mike and Liu, Yinhan and Goyal, Naman and Ghazvininejad, Marjan and Mohamed, Abdelrahman and Levy, Omer and Stoyanov, Ves and Zettlemoyer, Luke},
  journal={arXiv preprint arXiv:1910.13461},
  year={2019}
}

@article{thoppilan2022lamda,
  title={Lamda: Language models for dialog applications},
  author={Thoppilan, Romal and De Freitas, Daniel and Hall, Jamie and Shazeer, Noam and Kulshreshtha, Apoorv and Cheng, Heng-Tze and Jin, Alicia and Bos, Taylor and Baker, Leslie and Du, Yu and others},
  journal={arXiv preprint arXiv:2201.08239},
  year={2022}
}

@article{ouyang2022training,
  title={Training language models to follow instructions with human feedback},
  author={Ouyang, Long and Wu, Jeffrey and Jiang, Xu and Almeida, Diogo and Wainwright, Carroll and Mishkin, Pamela and Zhang, Chong and Agarwal, Sandhini and Slama, Katarina and Ray, Alex and others},
  journal={Advances in Neural Information Processing Systems},
  volume={35},
  pages={27730--27744},
  year={2022}
}

@article{shah2022flue,
  title={When flue meets flang: Benchmarks and large pre-trained language model for financial domain},
  author={Shah, Raj Sanjay and Chawla, Kunal and Eidnani, Dheeraj and Shah, Agam and Du, Wendi and Chava, Sudheer and Raman, Natraj and Smiley, Charese and Chen, Jiaao and Yang, Diyi},
  journal={arXiv preprint arXiv:2211.00083},
  year={2022}
}

@inproceedings{finred2022,
author = {Sharma, Soumya and Nayak, Tapas and Bose, Arusarka and Meena, Ajay Kumar and Dasgupta, Koustuv and Ganguly, Niloy and Goyal, Pawan},
title = {FinRED: A Dataset for Relation Extraction in Financial Domain},
year = {2022},
isbn = {9781450391306},
publisher = {Association for Computing Machinery},
address = {New York, NY, USA},
url = {https://doi.org/10.1145/3487553.3524637},
doi = {10.1145/3487553.3524637},
booktitle = {Companion Proceedings of the Web Conference 2022},
pages = {595–597},
numpages = {3},
location = {Virtual Event, Lyon, France},
series = {WWW '22}
}

@article{chen2021finqa,
  title={Finqa: A dataset of numerical reasoning over financial data},
  author={Chen, Zhiyu and Chen, Wenhu and Smiley, Charese and Shah, Sameena and Borova, Iana and Langdon, Dylan and Moussa, Reema and Beane, Matt and Huang, Ting-Hao and Routledge, Bryan and others},
  journal={arXiv preprint arXiv:2109.00122},
  year={2021}
}

@inproceedings{finread2021,
    title = "{F}in{R}ead: A Transfer Learning Based Tool to Assess Readability of Definitions of Financial Terms",
    author = "Ghosh, Sohom  and
      Sengupta, Shovon  and
      Naskar, Sudip  and
      Singh, Sunny Kumar",
    booktitle = "Proceedings of the 18th International Conference on Natural Language Processing (ICON)",
    month = dec,
    year = "2021",
    address = "National Institute of Technology Silchar, Silchar, India",
    publisher = "NLP Association of India (NLPAI)",
    url = "https://aclanthology.org/2021.icon-main.81",
    pages = "658--659"
    }

@article{zhang2023fingptrag,
  title={Enhancing Financial Sentiment Analysis via Retrieval Augmented Large Language Models},
  author={Zhang, Boyu and Yang, Hongyang and Zhou, Tianyu and Babar, Ali and Liu, Xiao-Yang},
 journal = {ACM International Conference on AI in Finance (ICAIF)},
  year={2023}
}

@article{grattafiori2024llama,
  title={The llama 3 herd of models},
  author={Grattafiori, Aaron and Dubey, Abhimanyu and Jauhri, Abhinav and Pandey, Abhinav and Kadian, Abhishek and Al-Dahle, Ahmad and Letman, Aiesha and Mathur, Akhil and Schelten, Alan and Vaughan, Alex and others},
  journal={arXiv preprint arXiv:2407.21783},
  year={2024}
}

@misc{openai_chatgpt,
  title        = {ChatGPT},
  author       = {OpenAI},
  year         = {2023},
  howpublished = {\url{https://chat.openai.com/}},
  note         = {Large language model accessed via ChatGPT interface}
}

@inproceedings{yang2020deep,
  title={Deep reinforcement learning for automated stock trading: An ensemble strategy},
  author={Yang, Hongyang and Liu, Xiao-Yang and Zhong, Shan and Walid, Anwar},
  booktitle={Proceedings of the first ACM international conference on AI in finance},
  pages={1--8},
  year={2020}
}

@article{yang2024finrobot,
  title={FinRobot: An Open-Source AI Agent Platform for Financial Applications using Large Language Models},
  author={Yang, Hongyang and Zhang, Boyu and Wang, Neng and Guo, Cheng and Zhang, Xiaoli and Lin, Likun and Wang, Junlin and Zhou, Tianyu and Guan, Mao and Zhang, Runjia and others},
  journal={arXiv preprint arXiv:2405.14767},
  year={2024}
}

@misc{jiang2023mistral7b,
      title={Mistral 7B}, 
      author={Albert Q. Jiang and Alexandre Sablayrolles and Arthur Mensch and Chris Bamford and Devendra Singh Chaplot and Diego de las Casas and Florian Bressand and Gianna Lengyel and Guillaume Lample and Lucile Saulnier and Lélio Renard Lavaud and Marie-Anne Lachaux and Pierre Stock and Teven Le Scao and Thibaut Lavril and Thomas Wang and Timothée Lacroix and William El Sayed},
      year={2023},
      eprint={2310.06825},
      archivePrefix={arXiv},
      primaryClass={cs.CL},
      url={https://arxiv.org/abs/2310.06825}, 
}

@article{team2023gemini,
  title={Gemini: a family of highly capable multimodal models},
  author={Team, Gemini and Anil, Rohan and Borgeaud, Sebastian and Alayrac, Jean-Baptiste and Yu, Jiahui and Soricut, Radu and Schalkwyk, Johan and Dai, Andrew M and Hauth, Anja and Millican, Katie and others},
  journal={arXiv preprint arXiv:2312.11805},
  year={2023}
}

@article{liu2024deepseek,
  title={Deepseek-v3 technical report},
  author={Liu, Aixin and Feng, Bei and Xue, Bing and Wang, Bingxuan and Wu, Bochao and Lu, Chengda and Zhao, Chenggang and Deng, Chengqi and Zhang, Chenyu and Ruan, Chong and others},
  journal={arXiv preprint arXiv:2412.19437},
  year={2024}
}

\end{document}